\documentclass[10pt,letterpaper]{article}
\usepackage[top=1.25in,left=0.75in,footskip=0.75in]{geometry}

\usepackage{amsmath,amssymb}

\usepackage{textcomp,marvosym}

\usepackage{cite}

\usepackage{nameref,hyperref}

\usepackage[right]{lineno}

\usepackage{microtype}
\DisableLigatures[f]{encoding = *, family = * }

\usepackage[table]{xcolor}

\usepackage{array}

\usepackage{graphicx}

\newcolumntype{+}{!{\vrule width 2pt}}

\newlength\savedwidth

\raggedright
\usepackage[aboveskip=1pt,labelfont=bf,labelsep=period,justification=raggedright,singlelinecheck=off]{caption}

\makeatletter
\renewcommand{\@biblabel}[1]{\quad#1.}
\makeatother

\date{}

\begin{document}
\vspace*{0.2in}

\begin{flushleft}
{\Large
\textbf\newline{Bounded Confidence under Preferential Flip:\newline  A Coupled Dynamics of Structural Balance and Opinions} 
}
\newline
\newline
\\
Antonio Parravano$^{1,2*}$,
Ascensi\'{o}n Andina-D\'{\i}az$^1$
Miguel A. Mel\'{e}ndez-Jim\'{e}nez$^1$
\\
\bigskip
\textbf{1} Dpto. Teor\'{\i}a e Historia Econ\'{o}mica, Universidad de M\'{a}laga, M\'{a}laga, Spain
\\
\textbf{2} Centro de Física Fundamental, Universidad de Los Andes, Mérida, Venezuela
\\
\bigskip

%
%





* parravan3@gmail.com

\end{flushleft}
\section*{Abstract}
In this work we study the coupled dynamics of social balance and opinion formation. We propose a model where agents form opinions under bounded confidence, but only considering the opinions of their friends. The signs of social ties -friendships and enmities- evolve seeking for social balance, taking into account how similar agents' opinions are. We consider both the case where opinions have one and two dimensions. We find that our dynamics produces the segregation of agents into two cliques, with the opinions of agents in one clique differing from those in the other. Depending on the level of bounded confidence, the dynamics can produce either consensus of opinions within each clique or the coexistence of several opinion clusters in a clique. For the uni-dimensional case, the opinions in one clique are all below the opinions in the other clique, hence defining a ``left clique'' and a ``right clique''. In the two-dimensional case, our numerical results suggest that the two cliques are separated by a hyperplane in the opinion space. We also show that the phenomenon of \textit{unidimensional opinions} identified by DeMarzo, Vayanos and Zwiebel (\textit{Q J Econ} 2003) extends partially to our dynamics. Finally, in the context of politics, we comment about the possible relation of our results to the fragmentation of an ideology and the emergence of new political parties.\newline





\section*{Introduction}
Dating back to Aristotle, who asserted long ago in his \textit{Politics}: ``Man is by nature a social animal'', it is generally accepted that human beings have a basic need to belong to a group where to share news and express their opinions. In our pursuit of belonging to a group, human beings are continuously forming new relationships and breaking old ones, which affects both the configuration of our social network and our way of thinking. Indeed, our opinions are often full of reminiscences of our friends and acquaintances' opinions. Not only this, our need to conform to other people’s opinions generally impulse us to establish new relationships with people we feel close to and to break an existing one when we fall apart. This phenomenon, known as homophily, has received the attention of sociologists, economists and physicists~\cite{Byrne71,Kandel78,Mark2003,ParravanoEA2007,CurrariniEA2009,CurrariniEA2010,FlacheMacy2011b,McPhersonEA2001}.

A stark example of human beings belonging to groups of like-minded information is politics. Casual observation shows that people not only tend to form connections with people with similar political preferences, but choose to receive information about politics and government from confirmatory-bias sources. A recent study by Pew Research Center confirms these ideas. The study looks at the ways people receive information about politics in three different settings: news media, social media and talk to friends and family. They conclude that ``\textit{Liberals and conservatives inhabit different worlds. There is little overlap in the news sources they turn to and trust. And whether discussing politics online or with friends, they are more likely than others to interact with like-minded individuals}''. (See ``Political Polarization and Media Habits'', Pew Research Center October 21, 2014.)

Despite the consensus that individual opinions and interpersonal relationships interact and coevolve, the research that studies coupled dynamics of both issues is still very scarce as compared to the abundant number of works that deal with these two issues in a separate way. (See the next section for a review of these works.) In this paper we propose a model that couples in a very natural way the dynamics of two well-known models of (i) evolution of social relationships (a structural balance model) and (ii) opinion formation (a bounded confidence model), and discuss some applications to politics. As far as we know, this is the first time that a dynamics of social balance and a dynamics of opinion formation are coupled to reduce both unbalanced social relations and opinion dissonance.

Regarding the evolution of the social structure, we base our dynamics on the
influential model of Antal et al~\cite{AntalEA2005}, one of the pioneering studies of
the dynamics of \textit{structural balance}. The theory of structural
balance, originary from Heider~\cite{Heider1946}, has a long tradition in sociology and
establishes that the evolution of friendships (positive links) and enmities
(negative links) are not just due to bilateral processes, but depend on the
relationships among triads of individuals. In particular, it postulates that
there exists a tension either (i) when three people are all enemies among
themselves, or (ii) when a person has two friends that are enemies between
themselves. In the first case, this tension only disappears (hence,
balancing the triad) when two of the enemies befriend and oppose to a common
``adversary''. In the second case, to restore balance, it is needed that
either the two enemies become friends (maybe due to the arbitration of the
common friend), or that one of the friendship relationships ends (so that,
afterwards, it holds that a pair of friends share a common enemy). This
theory was extended to account for the stability of networks in~\cite{CartwrightHarary1956}, obtaining that a complete graph is \textit{balanced} (i.e.,
free of tensions in all triads) if and only if either all individuals are
friends or there are two antagonistic
cliques, with all persons within each clique being friends and all pairs of
persons in different cliques being enemies (referred to as \textit{bipolar
state }by Antal et al~\cite{AntalEA2005}). In this respect, recently there has been high
efforts to compute the global level of balance of real world networks. (See~\cite{FacchettiEA2011}, which verifies that Epinions, Slashdots and WikiElections
-three large online social networks- are extremely balanced.)

Antal et al~\cite{AntalEA2005} propose two discrete-time dynamics of this theory, called
local triad dynamics and constrained triad dynamics. In these
models, at each time step, one (randomly selected) triad is revised. In the
unconstrained model (LTD), if the triad is unbalanced, one of it links is
randomly selected to switch its sign (i.e., either a friendship is broken or
two former enemies become friends) in order to balance the triad. In the
constrained model (CTD) the same holds, but the switches are only
implemented if they produce more balanced than unbalanced triads in the
network. In this respect, we base our network dynamics on the unconstrained formulation in~\cite{AntalEA2005}, but we depart from their (random) process of
selecting links in order to balance triads. In particular, we endow each
agent in the network with an opinion (which evolves over time, as described below) and
consider that, when a link must switch sign to stabilize a triad, it is the distance among the opinions of the
agents in the triad what determines which link flips. More precisely, we consider that when a friendship relationship is broken, it is between
the pair of agents with the highest difference of opinions whereas, when a
friendship relationship is formed, it is between the pair of agents with the
lowest difference of opinions. In this sense, we say that our model presents
\textit{preferential flip}, as compared to the random flip
assumption in~\cite{AntalEA2005}.

Regarding opinions, we consider that they evolve over time following Hegselmann and Krause~\cite{HegselmannKrause2002}, which is one of the most influential models of
opinion dynamics under bounded confidence. The idea of bounded confidence
refers to the fact that individuals update their opinions only taking into
account those individuals whose opinions differ from their own not more than
a certain confidence level. These models build on the idea that the
formation of opinions on complex social issues are typically subjected to a
confirmation bias~\cite{LordEA1979,LazarsfeldEA1948,LazarsfeldEA1954,Andina2007,Allahverdyan2014,BessiEA2015,DelVicarioEA2016a,DelVicarioEA2016b}. In particular, Hegselmann and Krause consider that
individuals account for a weighted average of the opinions of those
neighbors in the network that satisfy the confidence level, hence
generalizing the classical De Groot's model of opinion formation~\cite{DeGroot1974}. Due
to non-linearities, the model in~\cite{HegselmannKrause2002} is not
analytically solvable, so most of the results are based on simulations. By
applying Hegselmann and Krause's dynamics to our framework, we make
the evolution of the opinions depend on the network. In particular, when an
individual revises his opinion, he takes into account the average of the
opinions of those of his friends whose opinions differ from his own
not more than the confidence level, neglecting the opinions of his
enemies. We consider two cases: one in which agents have opinions on one issue, the other in which opinions contain two issues.

Because the dynamics we present in this work incorporates features of models of opinion formation under bounded confidence and of models of structural balance, we label our model BC-PF (bounded
confidence under preferential flip). The BC-PF model, in
addition to providing a sensible way to couple these two dynamics, offers
several contributions. First, it significantly reduces convergence time as
compared to Antal et al's~\cite{AntalEA2005} initial (LTD) formulation, without the need to
constrain the dynamics. In this sense, the BC-PF model provides a new way to reduce the time needed to reach social balance, maintaining a discrete-time social balance framework. Second, when the BC-PF dynamics leads to the bipolar state, it produces
segregation of agents in the opinion space, both when agents have opinions on a single issue and when opinions include two issues. Third, it shows that depending on the agents' confidence level, the system exhibits consensus of opinions within a clique or the coexistence of several opinion clusters, again both in the case with one and two issues. Fourth, we observe that the phenomenon of \textit{unidimensional opinions} identified by DeMarzo et al~\cite{DeMarzoEA2003} for dynamics leading to global consensus extends partially to our case. Last, the BC-PF dynamics may have applications to politics, potentially suggesting new arguments to explain the fragmentation of an ideology and the emergence of new political parties.

The remainder of the paper is organized as follows. The next section presents
a brief literature review. In the third section we present our model. The fourth and fifth sections present the results for the one-dimensional opinions and bi-dimensional opinions cases, respectively. In the sixth section we comment on the applicability of our results to politics. Finally, we conclude with a discussion of our results and possible extensions.

\section*{Related literature}

In this section we first present a brief review of models that study the
dynamics of social relationships, with a focus on those models dealing with
structural balance. Then, we review some models that study the formation of
opinions within fixed networks. Finally, we report on some recent papers
that consider the joint evolution of opinions and networks.

As aforementioned, the initial models of structural balance~\cite{Heider1946,CartwrightHarary1956} were static and simply aimed to identify the
stable (balanced) structures. Antal et al~\cite{AntalEA2005,AntalEA2006} propose a dynamics to
reach structural balance, a topic that has attracted a lot of attention from
the literature in the last decade. (See the survey in~\cite{ZhengEA2015}.) In this respect, given that many unbalanced graphs
(the jammed states) are stable points for the CTD (discrete-time)
dynamics in~\cite{AntalEA2005}, Ku\l akowski et al~\cite{KulakowskiEA2005} focus on a
continuous-time dynamics in order to identify a system that leads to
balanced networks from generic initial configurations. In their model,
subsequently investigated by~\cite{MarvelEA2011} and~\cite{SrinivasanEA2011}, they
consider the strength of the friendliness or unfriendliness (by adding a
weight to the sign of each link), and find that the system always converges
to a balanced state. As shown by~\cite{MarvelEA2011}, this state is the
all-friends configuration if the mean value of the initial friendliness
among the nodes is (strictly) positive, whereas it is a bipolar state with
two cliques of approximately equal size otherwise. (See also~\cite{TraagEA2013} for an alternative specification of the continuous-time dynamics.) Aguiar and Parravano~\cite{AguiarParravano2015} also extend Antal et al's model, aiming
to address the impact of partitioning the population into two groups, with
tolerant and intolerant individuals within each of them. They find that,
as the size of the system increases, two balanced solutions dominate:
segregation into two cliques and the isolation of intolerant agents.

On the other hand, the original models dealing with the evolution of opinions build
on French's theory of social power~\cite{French1956}, which was subsequently developed
in a more general form by Harary~\cite{Harary1959} and DeGroot~\cite{DeGroot1974}. The latter
considers that the individual's opinion at each period is the weighted
average of his neighbors' opinions in the previous period, a dynamics which
ultimately leads to consensus provided the network is strongly connected and
weak aperiodic. Many studies have analyzed variations and extensions of
these initial models. (See, for instance,~\cite{FriedkinEA1990,FriedkinEA1999,DeMarzoEA2003,Golub2010}.) Among them, the models
that are more closely related to our work are those of opinion formation
under bounded confidence, pioneered by Deffuant et al~\cite{DeffuantEA2000} and Hegselmann and Krause~\cite{HegselmannKrause2002}. The model in~\cite{DeffuantEA2000} mainly differs from that in~\cite{HegselmannKrause2002} in that, in the former, at each time step, individuals update their opinions by taking into account the opinion of a single
(randomly selected) neighbor. There have been a number of papers constructed
over these pioneering models exploring, for instance, the multi-dimensionality of
opinions~\cite{FortunatoEA2005,PluchinoEA2006,FlacheMacy2011} or heterogeneous confidence levels~\cite{KouEA2012,FuEA2015}. In this respect, see the surveys in~\cite{LorenzEA2007,CastellanoEA2009,AlbiEA2016}. Relatedly, Dandekar et al~\cite{DandekaraEA2013} and Groeber et al~\cite{GroeberEA2014} also study the dynamics of polarization of opinions, but they model the confirmation bias in a more continuous way than the bounded confidence models.

Recent studies consider the evolution of opinions on networks that, although
fixed, either are structurally balanced~\cite{Altafini2012,Altafini2013}, or rely on the theory of structural balance~\cite{HuEA2014}. Altafini~\cite{Altafini2012} considers a social community split into two antagonistic factions and suggests a class of dynamical systems as natural models for the dynamics of
opinion forming, considering that the influence of a friend is positive and
that of an adversary negative. (See~\cite{FlacheMacy2011,FanEA2012,XiaEA2016,ProskurnikovEA2016,MengEA2016,ShiEA2016} for other models that allow for negative influence.) In
this setup, Altafini~\cite{Altafini2013} finds that the process leads the opinions of all agents within each faction to be equal and to be exactly the opposite (same modulus but different sign) to the opinion in the other faction (which he calls bipartite consensus). Hu et al~\cite{HuEA2014}, relying on the theory of
structural balance, derive sufficient conditions for the consensus,
polarization or fragmentation behaviors of a multi-agent system under the
assumption that the signed network has a spanning tree. In a similar vein, M\"as and Flache~\cite{MasFlache2013} propose the so-called ``argument-communication theory of bi-polarization'', which
explains that initially homogeneous populations can segregate into groups
with opposed opinions, even in situations in which social influence is only
positive and individuals do not seek to distance from each other.

Some very recent papers analyze a similar context in models in which both
the network and the opinions vary over time, but the dynamics are not
coupled. In this respect, Proskurnikov et al~\cite{ProskurnikovEA2016} show that the results of bipartite consensus in~\cite{Altafini2012,Altafini2013} extend to the case of a
time-varying graph under some conditions regarding connectivity. (See also~\cite{MengEA2016}, which studies the consensus problem in networks with
antagonistic and cooperative interactions in a model where the network topology graph may vary over time.) Xia et al~\cite{XiaEA2016} focus on structurally
unbalanced networks. They first show that, if the network includes a strongly connected subnetwork (containing negative links) that is
structurally balanced, then the agents of the subnetwork polarize into two
opinions and the opinions of all other agents in the network spread between
these two opinions, a result that they extend to time-varying networks.

In the same spirit of our work, there is a strand of the literature that
considers the joint evolution of opinions and networks, but where the
network mainly evolves to reduce the dissonance of opinions between agents,
rather than relying on structural balance arguments. In this respect, the
seminal paper by Holmes and Newman~\cite{HolmeEA2006} consider that agents modify
the network by removing and rewiring links towards agents with the same (discrete) opinion than theirs, and they revise opinions mimicking the opinion of a randomly selected
neighbor. They find that there is a phase transition (in the parameter controlling the balance of links and opinions updates) from a regime with
many opinions clusters to one close to global consensus. Some papers have
subsequently explored related models based on a mechanism of ``removing and rewiring'' links that
increases the opinion concordance, mainly dealing with discrete opinions~\cite{FuWang2008,NardiniEA2008,VazquezAE2008,IniguezEA2009,DurrettEA2012,YiEA2013,ArifovicEA2015}. With a similar
mechanism, some models study the co-evolution using a bounded confidence criteria to
update opinions~\cite{KozmaBarrat2008,DelVicarioEA2016a}.

Other (coevolutionary) models do not consider a rewiring mechanism, but rather assume that
the initial network is complete and links can only be removed throughout
the dynamics~\cite{GilZanette2006,ZanetteGil2006}. Differently, Ehrhardt et al~\cite{EhrhardtEA2006,EhrhardtEA2008} study the case in which links are
created over time depending on agents opinions (or attributes) and, at the
same time, all links in the network are taken to disappear with positive probability along the dynamics. (See also the related literature in economics dealing with the joint evolution of conventions and networks~\cite{JacksonWatts2002,GoyalVega2005,FeriMelendez2013,HellmanStaudigl2014}.) Additionally, Benczih et al~\cite{BenczikEA2009} study the case where each agent updates all her links simultaneously. Finally, Flache and Macy~\cite{FlacheMacy2011} allow for
the presence of negative links and study the co-evolution of node opinions
and link weights in a variety of small-world networks. In this study, agents'
opinions are updated to the weighted average of their neighbors, and link
weights increase (or decrease) in proportion to the opinion concordance (or
discordance) in the various opinion issues under consideration.

In the model proposed
here, instead of assuming that the social links evolve to reduce the opinion
discordance between pairs of agents (as in the majority of coevolutionary models discussed so far), we assume that unbalanced structures
(specifically triads of agents) evolve to balanced structures choosing the
option that reduce the opinion discordance the most. This approach therefore
allows us to examine the interplay between the stress in three agents'
conflicting relations and the stress in two agents' opinion dissonance.

\section*{The model}
\label{sec:model}

We consider a process of opinion formation in a fully connected network with $N$ social agents. We assume that each agent in the network is characterized by a vector $\overrightarrow{x_i}(t)$, that
represents the opinion of agent $i$ at time $t$ on the $F$ issues at stake. For each issue $f$, $x_{i,f}(t)\in[0,1]$.

We interpret agents as nodes in a graph and a social relationship between two agents $i$ and $j$, $s_{ij}$, as a link between the two agents, where links can be either positive or negative. We denote
by $s_{ij}(t)=+1$ a situation where, at time $t$, agents $i$ and $j$ in the network are friends $(+)$, and by $s_{ij}(t)=-1$ a situation where they are enemies $(-)$.

We assume that interactions among individuals in the network occurs in triads. This is a classical feature of Heider's balance theory~\cite{Heider1946}. A triad $ijk$ is said to be balanced if $s_{ij}\times s_{ik}\times s_{jk}=1$ and unbalanced if $s_{ij}\times s_{ik}\times s_{jk}=-1$. A balanced triad necessarily embeds either (i) three agents who are all friends of each other, or (ii) one friendship relationship and two enmity ones, i.e., two friends with a common enemy. Hereafter, we refer to balanced triads of types (i) and (ii) as $(+++)$ and $(+--)$ triads, respectively. As for unbalanced triads, there are also two types: one in which the three agents are all enemies of each other, refereed to as a triad of type $(---)$; the other with one enmity relationship and two friendship ones, i.e., two enemies with a common friend, refereed to as a triad of type $(-++)$.

In the present study we investigate the coupled evolution of both social balance and opinions in the network.

To study the dynamics of social balance, we follow Antal et al~\cite{AntalEA2005} and assume that every time step $t$ a triad is randomly chosen to be revised. If the triad is balanced, nothing occurs. Unbalanced triads of type $(---)$ stabilize by flipping one link to positive to reach a balanced configuration of type $(+--)$. Unbalanced triads of type $(-++)$ have two ways to stabilize: With probability $p\in(0,1)$ the negative link flips to positive to reach the balanced configuration $(+++)$, and with probability $1-p$ one of the positive links flips to negative to reach a balanced configuration of type $(+--)$. Antal et al assume that the link that flips to reach balance is chosen at random (with uniform probability). As described below, we depart from this (random flip) formulation. Fig~1 (left and central panels) represents the transition in~\cite{AntalEA2005}.

To study the dynamics of opinions in the network, we follow Hegselmann and Krause's~\cite{HegselmannKrause2002} model of opinion formation with Bounded Confidence (hereafter BC).
The simplest version of this model considers that agent $i$ adjusts his opinion at time $t+1$ by taking the average (including the focal agent $i$)
of the opinions of those agents $j$ whose opinions differ from his own not more than a confidence level $\varepsilon\in[0,1]$.

\begin{equation}
x_i(t+1)=\frac{1}{\eta_i} \sum_{j \in \{\nu_i\}(t)}x_j(t)
\end{equation}
where $\{\nu_i\}(t)=\{j\in N : |x_j-x_i| \leq \varepsilon\}\cup \{i\}$, and $\eta_i$ is the number of agents
in $\{\nu_i\}(t)$.

Because our model integrates two models that haven't talked to each other before, there is a need to accommodate certain aspects of previous models. In the present paper we assume:

First, that  an agent's opinions are never influenced by his enemies. In other words, we consider that agent $i$'s opinion is only influenced by those agents $j$ for which both conditions $s_{i,j}=+1$ and $|x_j-x_i| \leq \varepsilon$, hold.

\begin{figure}[!ht]
\hbox{\includegraphics[scale=1.0,angle=0]{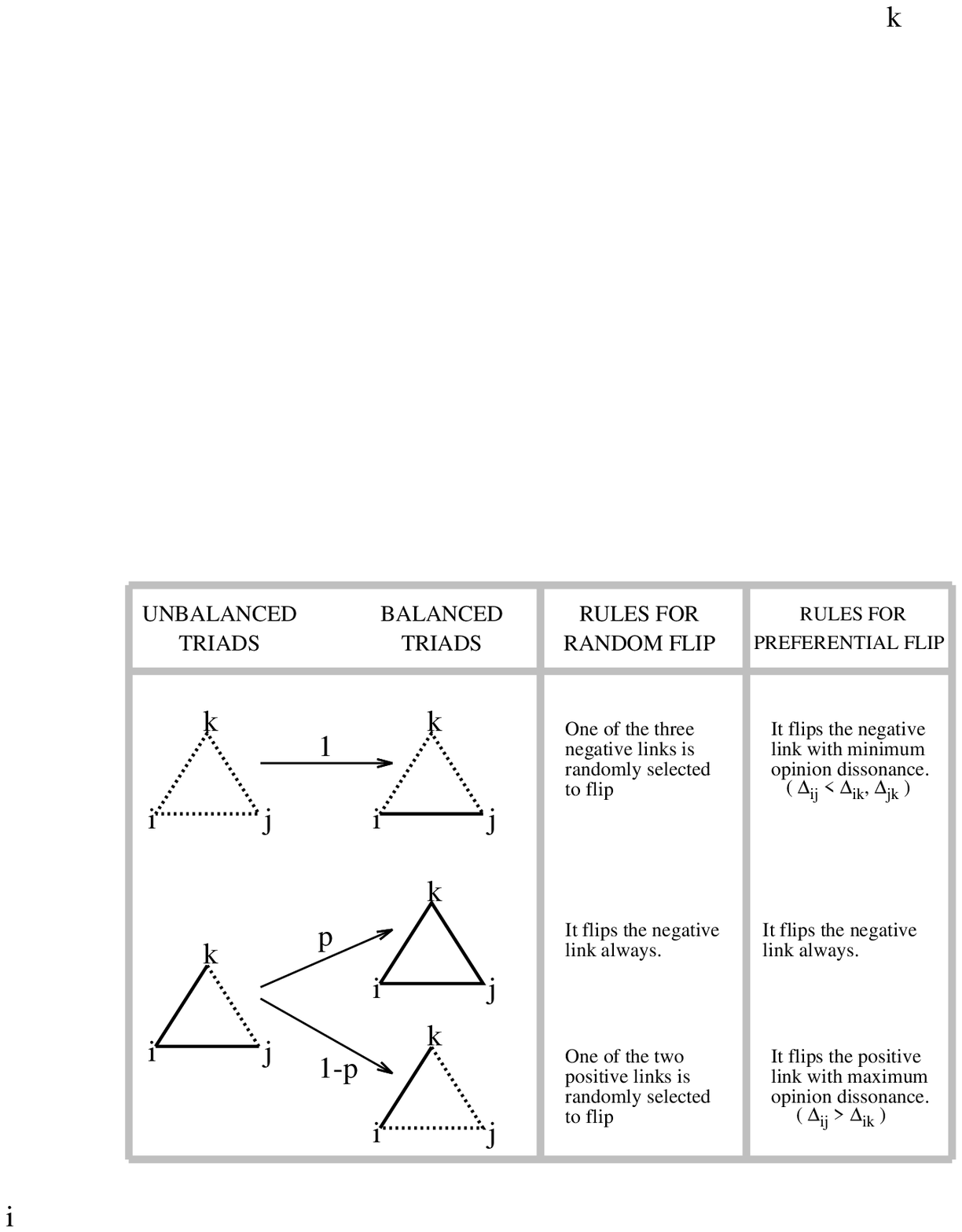}}
\caption{\small Left panel: Representation of balanced an unbalanced triads, where solid and dotted lines represent friendly $(+)$ and enmity $(-)$ relationships, respectively. Central panel: Rules to flip from an unbalanced to a balanced triad in Antal et al~\cite{AntalEA2005}. Right panel: Rules to flip from an unbalanced to a balanced triad in the BC-PF model.}
\label{rules}
\end{figure}

Second, that whenever an unbalanced trial is chosen to be revised, the link that flips is not randomly chosen but depends on opinion differences $\Delta_{ij}=|x_i-x_j|$. In particular, we
assume that in a triad of type $(---)$, the link that flips to positive is the one with the smallest $\Delta$.  Similarly, in a $(-++)$ triad, the positive link that flips to negative (with 
probability $1-p$) is the one with the largest $\Delta$. (With probability $p$, the negative link flips to positive.) These rules are summarized in Fig~\ref{rules} (left and right panels). Note that 
these assumptions affect the dynamics of social balance and, because the latter changes the signs of relationships, it also affects the dynamics of opinions in the network.

We term the balancing mechanism considered in this paper as Bounded Confidence under Preferential Flip (BC-PF).

Finally, note that in a system of $N$ agents, there are  $N \times (N-1) \times (N-2)/6$ triads and $N$ agent's opinions. For the results reported here, we consider the case in which triads and opinions are revised evenly, i.e, we update (at random) an agent's opinion every $(N-1) \times (N-2)/6$ triads updates, which are also updated at random. In this line, we define a \textit{round} as $N \times (N-1) \times (N-2)/6$ time steps. That is, in average, in one round all triads and opinions are revised.

\section*{One-dimensional opinions}

\subsection*{Social balanced configurations}

We start by comparing the social balanced configurations that the present BC-PF model produces with the ones in Antal et al~\cite{AntalEA2005}.
In the first part of their paper, Antal et al show that for the Local Triad Dynamics (LTD) model, if $p\geq 1/2$ the all-friends configuration is reached in the absorbing balanced state; whereas if $p<1/2$ the network evolves to a bipolar balanced state with two cliques. The BC-PF mechanism presents a similar behavior. More precisely, we obtain that the BC-PF model converges to a bipolar state (a two clique configuration with $s_{ij}=+1$ for any $i,j$ in clique $C\in\{C_1,C_2\}$, and $s_{ij}=-1$ for any $i\in C_1$ and $j\in C_2$) for $p<1/2$, and to the all-friends configuration (with $s_{ij}=+1$ for any $i,j$) when $p>1/2$. As it will be shown next, the main difference between the LTD and BC-PF model is that the latter substantially reduces the time to reach social balance and that it can produce the segregation of agents into opinion groups that coexist within each clique.

In the following we restrict our study to cases with $p<1/2$. Note that for $p\geq 1/2$, once the all-friends configuration is reached, our model reduces to the classical opinion formation model.

\subsection*{Convergence time}

Let us denote by $T_\text{typ}$ the typical time to reach social balance. Antal et al show that for the LTD model, the number of iterations required to reach social balance grows
super-exponentially with the size of the system ($T_\text{typ} \varpropto exp(N^2)$ for $p<1/2$). Then, in order to reduce the convergence time,
they introduce a restriction on the LTD dynamics that consists in allowing a link to flip only if the total number of unbalanced triads after the flip does not increase. They refer to this balancing mechanism as Constrained Triad Dynamics (CTD).

The CTD has two interesting effects on the evolution of the network structure. On the one hand, it substantially reduces the typical time to reach the absorbing state (the CTD dynamics converge to a
balanced structure in a time that scales as $\ln N$). On the other hand, as a side effect, the system can be trapped in jammed states, i.e., a structure with unbalanced triads in which no flip
increases the number of balanced triads.

The problem with CTD, as argued in~\cite{Abel2009}, is that it assumes agents with extraordinary calculative
capabilities, who are able to compute the effect of a link flip in a triad on all other triads. This aspect is particularly difficult to justify when networks are relatively large.

\begin{figure}[!ht]
\hbox{\includegraphics[scale=1.0,angle=0]{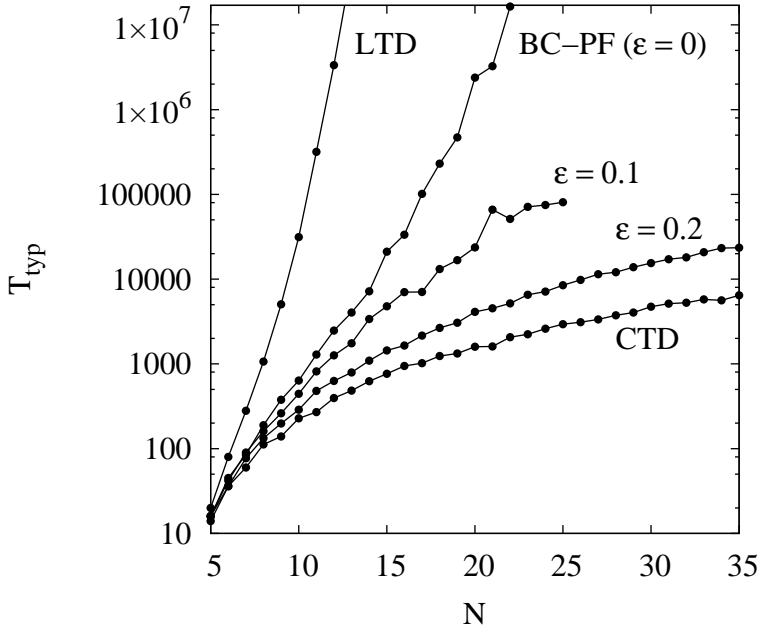}}
\caption{\small Typical (median) number of iterations $T_\text{typ}$ -in time steps- to converge to a balanced configuration as a function of the number $N$ of agents in the system
for $p=0.1$. The upper and lower curves correspond to the pure Antal (LTD) and the Constrained Triad Dynamics (CTD) models, respectively. The other three curves correspond to the BC-PF
dynamics with $\varepsilon=0$, $\varepsilon=0.1$  and $\varepsilon=0.2$, respectively. Each point represents the median value of 1000 different runs departing from random
initial conditions with equal proportion of positive and negative links.}
\label{typical-time}
\end{figure}

Interestingly, the BC-PF dynamics has also the property to significantly reduce the convergence time compared to the LTD dynamics. This is illustrated in
Fig~\ref{typical-time}, which presents the typical time $T_\text{typ}$ to reach social balance as a function of the size of the system $N$, for the LTD and the CTD dynamics proposed by Antal et al, and for the BC-PF dynamics. For the BC-PF we present results for the cases $\varepsilon=0,0.1$ and $0.2$. The case with $\varepsilon=0$ refers to a situation where agents do not change
their opinions over time (as they only look at friends with exactly their same opinion). Note that even though initial opinions do not change in this case in time, the PF mechanism has an important
effect on the dynamics of the network, considerably reducing the time of convergence compared to the LTD case. (The dependence becomes in this case closer to $\ln T_\text{typ} \propto N$.)  As $\varepsilon$ increases, the curvature of $\ln T_\text{typ}(N)$ decreases, becoming negative at some value of $\varepsilon$ between 0 and 0.1, and reaching a lower limit for $\varepsilon \sim 0.25$. As Fig~\ref{typical-time} shows, when $\varepsilon=0.2$ the scaling is close to the one in
the CTD case. We also observe that $T_\text{typ}$ decreases with $p$, but the dependence is weaker than with $\varepsilon$.

\subsection*{Segregation of opinions}

\begin{figure}[!ht]
\hbox{\includegraphics[scale=1.0,angle=0]{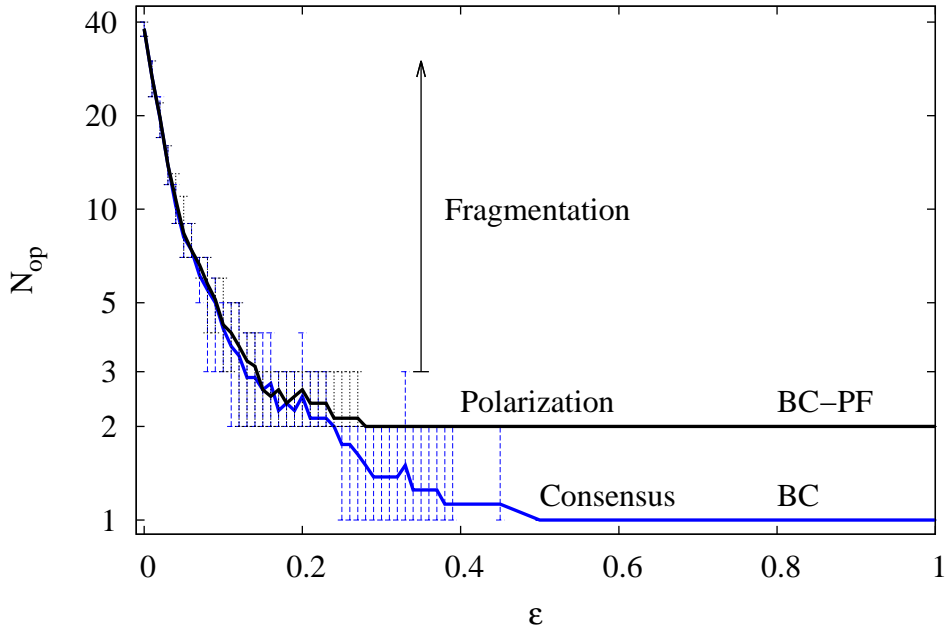}}
\caption{\small Number of different opinions $N_\text{op}$ in final balanced configurations as a function of the bounded confidence parameter $\varepsilon$ for 40 agents and  $p=0.1$. The grey curve
corresponds to the BC model and the black curve to the BC-PF model. The curves represent the mean value of $N_\text{op}$ in 8 simulations with random initial
opinions. The initial links are also set at random with equal probability for friendship (+1) and for enmity (-1). The vertical bars indicate the maximum and minimum values of $N_\text{op}$ in the 8 simulations.}
\label{Nop}
\end{figure}

\begin{figure}[!ht]
\hbox{\includegraphics[scale=1.0,angle=0]{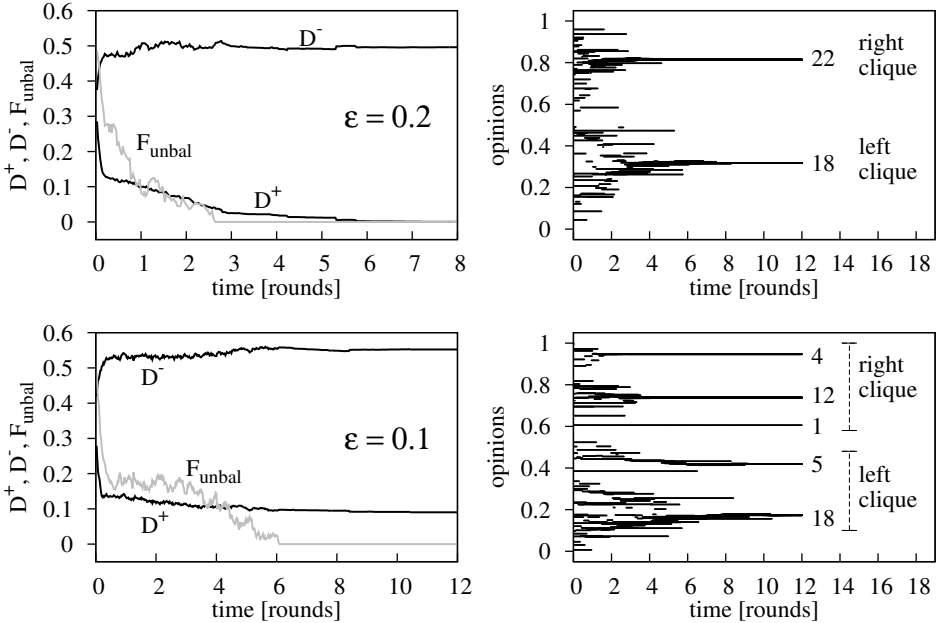}}
\caption{\small Evolution for a 40 agents system with $p=0.1$ and either $\varepsilon=0.2$ (upper panels) or $\varepsilon=0.1$ (bottom panels).
The two cases shown correspond to a particular run converging to social balance in few rounds; $\sim 3$ rounds in the case with $\varepsilon=0.2$ and
$\sim 6$ rounds in the case with $\varepsilon=0.1$. Time is given in round units. Left panels show the evolution of the intra-clique opinion diversity $D^+$, the inter-clique opinion separation $D^-$ and the fraction of unbalanced triads $F_\text{unbal}$. Right panels show the corresponding evolution of the opinions. The labels indicate the number of agents having the same final opinion.
In the case with $\varepsilon=0.2$ all the agents in a clique share the same opinion (intra-clique consensus $D^+=0$). In the case with $\varepsilon=0.1$ opinions converge to 5 different values in $\sim 10$ rounds. Agents in the ``left clique'' are grouped into two opinion clusters (of $18$ and $5$ agents), and agents in the ``right clique'' are grouped into three clusters (of $1$, $12$ and $4$ agents). Within each cluster there is consensus of opinions.}
\label{f-evo-40}
\end{figure}

We observe that the BC-PF dynamics yields both a segregation of agents into two cliques and a segregation of opinions. Simulations show that
when the
system reaches the balanced state, the number of different opinions in the system depends on the confidence interval. Fig~\ref{Nop} represents the number of different opinions in the
system $N_\text{op}$ as a function of the confidence interval $\varepsilon$, for the BC model in Hegselmann Krause~\cite{HegselmannKrause2002} and the BC-PF model. We observe that the BC model has a critical
value, $\varepsilon_c \simeq 1/2$, above which the system \textit{always} reaches consensus (one opinion). In contrast to this, the BC-PF dynamics exhibits consensus of opinions (within each clique) for $\varepsilon \gtrsim 0.25$. Thus, as compared to the BC model, the BC-PF dynamics lightens the requirement on the agent's confidence interval for consensus to be reached. To have an intuition for this result, note that in the balanced state of the BC-PF model there is a segregation of agents into two cliques according to their opinions (the low opinion values in one clique and the high ones in the other clique; see Fig \ref{f-evo-40}). Because in this case no agent has his opinion affected by the opinion of an agent who is not in the same clique, a confidence interval of $\varepsilon\sim 0.25$ is already enough to make, at the clique level, that extreme opinions get influenced by moderate ones, which guarantees intra-clique convergence. Note also that because in the BC-PF model consensus can only occur at the clique level, the present dynamics produces polarization (two cliques and two opinions in the system). Interestingly,
the BC-PF model suggests that segregation of agents into cliques
and polarization of opinions is the only balanced configuration for fairly large enough confidence intervals (namely,
$\varepsilon \gtrsim 0.25$). For $\varepsilon <0.25$, the number of opinions in the system can be relatively large, so as the number of opinions within a clique.

In order to better understand whether segregation of agents into cliques is related to segregation of opinions, we introduce the
following two measures. We define $D^+$ as the intra-clique opinion diversity, or friends' opinion dispersion,

\begin{equation}
D^+=\frac{1}{n^+} \sum_{i,j} \delta(s_{ij}-1) |x_j-x_i|
\end{equation}

\noindent and $D^-$ as the inter-clique opinion separation, or enemies' opinion divergence,

\begin{equation}
D^-=\frac{1}{n^-} \sum_{i,j} \delta(s_{ij}+1) |x_j-x_i| \,
\end{equation}

\noindent where $\delta(x)=1$ when $x=0$ and $\delta(x)=0$ otherwise, whereas $n^+$ and $n^-$ are respectively the number of positive and negative links.

We observe that the BC-PF dynamics always produces a neat differentiation of opinions between cliques. This is clear in the left hand side panels of Fig~\ref{f-evo-40}, that represent
the intra-clique (friends') opinion diversity $D^+$, the inter-clique (enemies')
opinion separation $D^-$ and the fraction of unbalanced triads $F_\text{unbal}$ as a function of time. Simulations
show that, for a relatively small number of rounds $t$, the system reaches a balanced state with two cliques, where agents in each clique share similar opinions ($D^+\simeq 0$); and agents of
different cliques end up having substantially different points of view of the issue at stake ($D^-\simeq 1/2$).

\begin{figure}[!ht]
\hbox{\includegraphics[scale=1.0,angle=0]{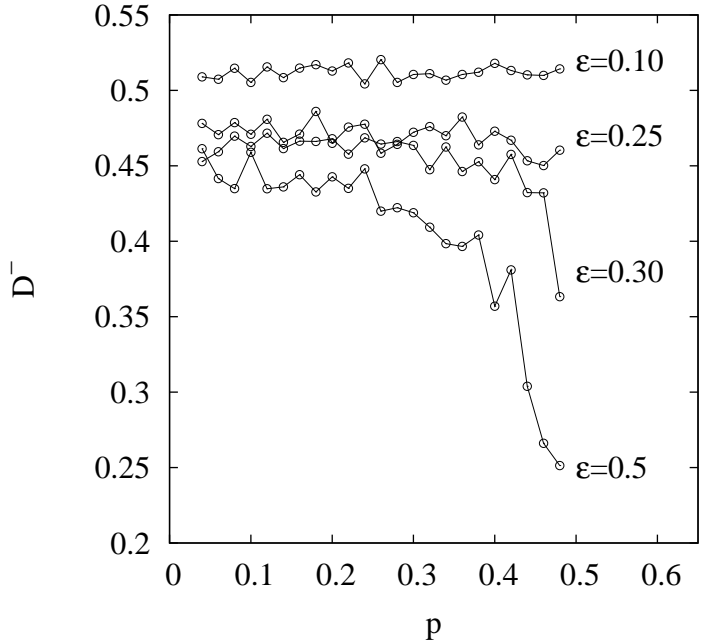}}
\caption{\small Average of $D^-$ in 100 runs as a function of parameters $\varepsilon$ and $p$ for a system of 40 agents.}
\label{Dm-e-p}
\end{figure}

As suggested by Fig~\ref{f-evo-40}, the BC-PF model produces inter-clique opinion separation that stabilizes around one half. In Fig~\ref{Dm-e-p} we show that this result is quite robust to changes in the confidence interval $\varepsilon$ and in the probability $p$. In fact, we observe that except for values of $p$ and $\varepsilon$ close enough to $1/2$,  inter-clique separation $D^-$ is $\sim 1/2$. To have an intuition for this result, first note that the preferential flip mechanism produces the segregation of agents into cliques according to their opinions. Hence, in the balanced configuration, opinions in one clique are all below (or above) opinions in the other clique -as explained below. Roughly, if $\varepsilon$ and $p$ are not very large, the average opinion is $\sim1/4$ ($\sim3/4$) in the lower (upper) clique, yielding an inter-clique opinion separation around $1/2$. When $\varepsilon$ and $p$ approach $1/2$, at the initial stages of the dynamics agents get influenced by a broader set of neighbors, which results in a smaller inter-clique opinion separation at the absorbing state.

Regarding the intra-clique opinion diversity, a closer look at the upper-left panel of Fig~\ref{f-evo-40}, corresponding to $\varepsilon=0.2$,
shows that convergence of opinions within cliques is reached ($D^+=0$); whereas the lower-left panel, corresponding to
the case $\varepsilon=0.1$, shows that more than one opinion coexist within each clique in the balanced configuration ($D^+\simeq 0.1$). To better illustrate this point,
we present the right hand side panels of Fig~\ref{f-evo-40}, that represents the confluence of opinions in groups as time goes on.
We observe that in all cases the number of opinions in a clique reduce with time. However, as already mentioned, the bottom panels of
Fig~\ref{f-evo-40} illustrate a situation in which $\varepsilon$ is relatively small and consensus within cliques is not reached.
Interestingly, we observe that in this case, the BC-PF dynamics produces the segregation of agents into two cliques according
to their opinions, where opinions in one clique are all below (or above) the opinions in the other clique.
This perfect segregation of opinions in cliques allows us to talk about the ``left clique'' and the ``right clique'', as the clique
containing the agents with the lower and higher opinion values, respectively. We observe that this is a quite general result, as starting from
random initial conditions we have never reached a balanced configuration where opinions are randomly distributed between cliques. (Note that this is usually
the case, except for very particular initial conditions, in which case different cliques may contain similar opinion values. For example,
when the stating network is already balanced but opinions within cliques are random.) Indeed, this segregation also holds in the extreme case
of $\varepsilon =0$, where even though agents' opinions do not evolve, agents segregate in two cliques in such a way that the left clique
includes the agents with the lower opinion values and the right one the agents with the higher opinion values.

We have also run simulations in which, departing from the same initial conditions as those used in Fig~\ref{f-evo-40}, we progressively double the speed at which opinions are updated up to the point that we update both one randomly selected triad and one randomly selected opinion at each iteration. We do not observe any substantial change in the behavior of the system with respect to the cases reported in Fig~\ref{f-evo-40}, except for a tendency of the typical convergence time to decrease as the speed of opinion update increases. In particular, we observe that the number of final opinions and their positions tend to remain unaltered. Only in some exceptional cases one opinion arises or disappears but, in any case, we find that for the range of speeds explored, the final values $D^+$ and $D^-$ are always very close to the corresponding counterparts depicted in Fig~\ref{f-evo-40}.

\subsection*{Size of cliques}

Antal et al show in their Fig~5 an unexpected feature of the Constrained Triad Dynamics model. They obtain that, in contrast to the LTD model, when the proportion of positive links in the initial state ($\rho_0$) increases, the CTD model exhibits a phase transition at $\rho_0\sim 0.65$ from two final cliques of nearly the same size to the all-friends configuration.

\begin{figure}[!ht]
\hbox{\includegraphics[scale=1.0,angle=0]{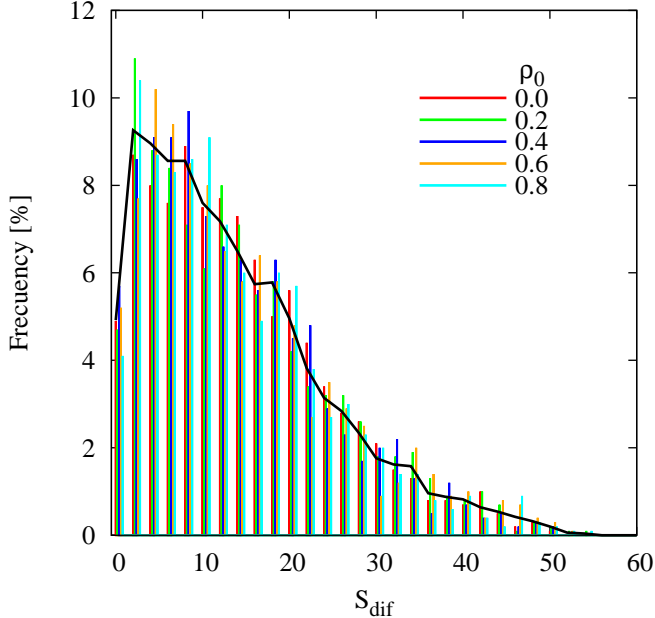}}
\caption{\small Distribution of size differences $S_\text{dif}$ -in per cent points- between the two cliques in a system with $N=100$, $p=0.3$
and $\varepsilon=0.2$. Each colored vertical line represents the frequency of clique size differences for a different proportion of
positive links $\rho_0$ in the initial conditions, as indicated in the legend. The back curve represents the mean of the previous frequencies. We run 1000 simulations for each $\rho_0$ value. Each vertical line represents the
frequency in 2\% bins and the different cases has been shifted each other to facilitate visualization.}
\label{dif-puntos}
\end{figure}

The BC-PF model differs from the CTD model in this respect. More precisely, we obtain that the BC-PF dynamics does not present a transition, as for any $\rho_0\in[0,1)$ and $\varepsilon\in[0,1]$, if $p<1/2$, the system always converges to a balanced configuration with two cliques. This aspect is clear in Fig~\ref{dif-puntos}, that shows the distribution of sizes differences $S_\text{dif}$ -in percent points- between the two cliques in the BC-PF model. We define $S_\text{dif}=\frac{|C_1-C_2|}{N}100$, with $C_1$ and $C_2$ being the size of cliques 1 and 2, respectively. We observe that the distribution of size differences $S_\text{dif}$ does not vary with the proportion of positive links in the initial state $\rho_0$. In this sense, the BC-PF model is robust to changes in $\rho_0$.

\section*{Multi-dimensional opinions}

In this section we extend the one-dimensional case to the case where agents have an opinion on more than one issue. Let $\overrightarrow{x_i}(t)$ be the vector of opinions of agent $i$ at time $t$ on the $F$ issues at stake. We define the distance between the vector of opinions of agents $i$ and $j$ as:
\begin{equation}
\Delta_{ij}(t)=\max\{\, |x_{i,f}(t)-x_{j,f}(t)| \, \text{for} \, f=1,\cdots,F\, \}
\end{equation}
where $x_{i,f}(t)\in[0,1]$ denotes agent $i$'s opinion on the \textit{f}-th issue at time $t$. Our definition of $\Delta_{ij}$ implies that two agents have opinions closer than $\varepsilon$ when they
differ in less than $\varepsilon$ in all the $F$ issues under consideration. This definition of $\Delta_{ij}$ is included in the opinion
dynamics, as well as in the network dynamics in the same way than in the one-dimensional case discussed before. In particular:\smallskip

1) $\Delta_{ij}(t)$ is used to determine the set of agents $\{\nu_i\}(t)$ inside the confidence level of agent $i$. That is, those agents $j$ with a link to $i$ that satisfies both the confidence boundary condition $\Delta_{ij} \leq \varepsilon$ and the friendship condition $s_{ij}=+1$. For each issue $f$ in $\overrightarrow{x_i}$, agent $i$'s opinion on this issue $x_{i,f}$ is updated following the Hegselmann Krause (2002) rule, that considers the opinions on the \textit{k}-th issue of all the agents in $\{\nu_i\}(t)$.\smallskip

2) $\Delta_{ij}(t)$ is used to determine the evolution of unstable triads. Thus, analogously to the one-dimensional case, we assume that in a triad of type $(---)$, the link that flips to positive is the one with the smallest $\Delta$. Similarly, in a $(-++)$ triad, the positive link that flips to negative (with probability $1-p$) is the one with the largest $\Delta$.\smallskip

Alternative definitions for $\Delta_{ij}$, as for example the geometrical distance $|\overrightarrow{x_i}(t)-\overrightarrow{x_j}(t)|$, yield similar results.

Before going into the analysis of the dynamics of social balance and opinions in the multi-dimensional case, it is worth mentioning some general ideas that we observe in this case. First, that if at some time $t$ (for example at $t=0$) all the agents' opinions in one issue are identical, say $x_{i,f}=x_0$ for issue $f$ and agent $i=1,\cdots,N$, then the opinions of the $N$ agents on issue $f$ will remain unchanged. This special case shows that our model can generate polarization in some issues and consensus in others. Second, that in a balanced configuration, it could occur that two agents in different cliques have similar opinions (closer than $\varepsilon$) on some issues. That is, that enemies share similar points of view on certain issues. This is due to our assumption that agents only consider the opinions of friends in their updating process. This feature could be modified by adding a term that produces the repulsion of enemies' opinions. (See the Discussion section.)

\subsection*{Bi-dimensional opinions}

\begin{figure}[!ht]
\hbox{\includegraphics[scale=1.0,angle=0]{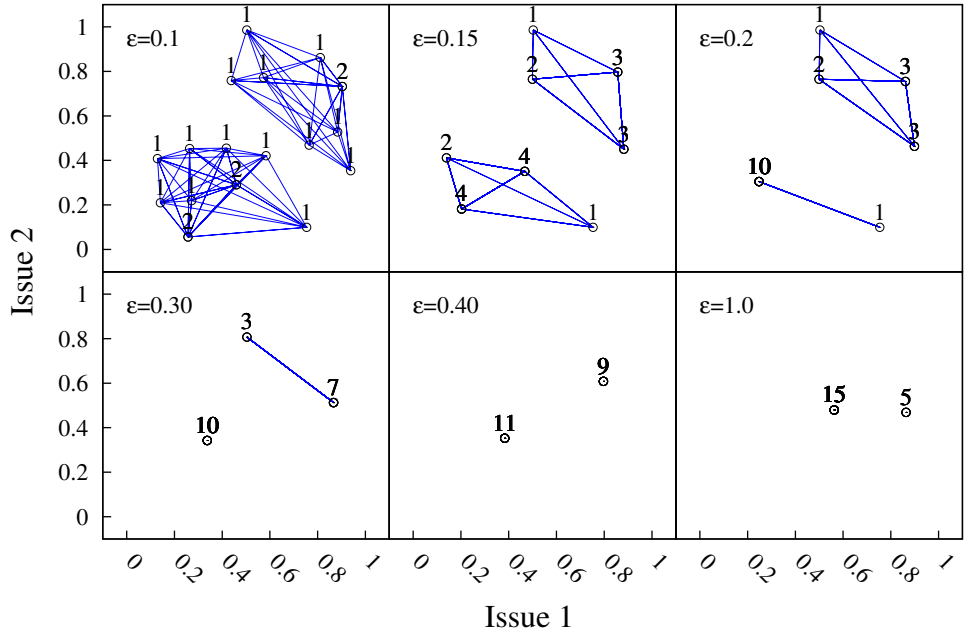}}
\caption{\small Six final absorbing states for a 20 agents system, starting from same initial condition (but different confidence levels $\varepsilon$, indicated at the upper-left corner of each panel). The initial opinions are shown in the upper-left panel of Fig~\ref{evo-e0p20}. The random initial network has the same number of positive and negative links, and we take $p=0.3$. The numbers in each panel indicate the number of agents with the same opinion (the number of agents in an opinion cluster). The grey lines show the positive links between agents. In the absorbing states there are neither negative links between agents in an opinion cluster nor negative links between agents in a clique. In the cases with $\varepsilon=0.4$ and $\varepsilon=1$, there is consensus of opinions in each clique.}
\label{panel-varia-e}
\end{figure}

\begin{figure}[!ht]
\hbox{\includegraphics[scale=1.0,angle=0]{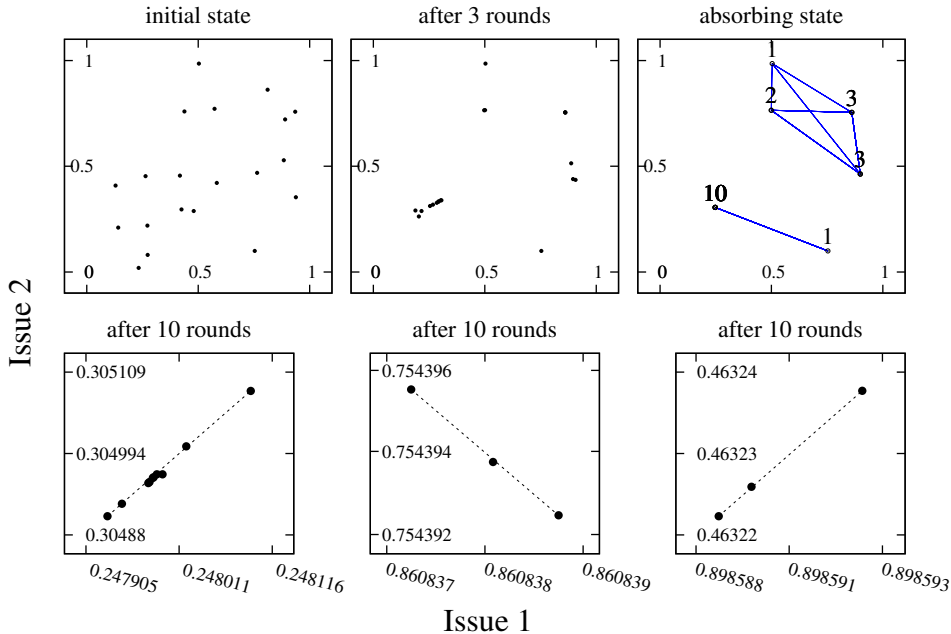}}
\caption{\small The upper panels present three snapshots of the evolution of opinions in a system for 20 agents, $\varepsilon=0.2$, $p=0.3$, and with a distribution of initial opinions that is shown in the upper left panel. The final absorbing state (shown in the upper right panel) corresponds also to the the upper right panel in Fig~\ref{panel-varia-e} (case $\varepsilon=0.2$). The three bottom panels show a zoom of the distribution of opinions of the members of an opinion group  just before reaching the final absorbing state. The bottom left, center and right panels correspond to the groups with 10, 3, and 3 agents, respectively.}
\label{evo-e0p20}
\end{figure}

We next investigate the coupled evolution of social balance and opinions in the case of two issues ($F=2$). Fig~\ref{panel-varia-e} illustrates the final configuration in the BC-PF dynamics with 
bi-dimensional opinions for different values of the confidence parameter $\varepsilon$. Each panel shows the position of every agent $i$'s opinions in the plane $x_{i,1}$ vs. $x_{i,2}$ in the final 
absorbing state, for the values of $\varepsilon$ quoted in the upper-left corner of each panel. Initial conditions are the same in the six panels. That is, we consider same initial link signs and same 
opinion values (shown in the upper-left panel of Fig~\ref{evo-e0p20}).

The analysis of this case yields a number of interesting analogies to the one-dimensional case, suggesting the robustness of these results. In particular, we observe the following analogies.

First, we observe that the BC-PF model with bi-dimensional opinions yields both a segregation of agents into two cliques and a segregation of opinions. This is clear in Fig~\ref{panel-varia-e}, where positive links are represented by grey links and the absence of a line indicates a negative link.

Second, it is usually the case that several agents converge to the same opinion and form an opinion cluster. In line with the one-dimensional case, we observe that the number of opinion clusters is inversely related to the confidence level $\varepsilon$. Thus, in the case where agents look at all their friends, i.e., $\varepsilon=1$, each clique exhibits consensus of opinions. This is also the case for $\varepsilon$ sufficiently large (see, for instance, the case $\varepsilon=0.4$ in Fig~\ref{panel-varia-e}). In contrast, when the confidence interval is small enough, we observe several opinion clusters within each clique (see the cases $\varepsilon=0.1$, $0.15$, $0.2$ and $0.3$ in Fig~\ref{panel-varia-e}). The number of agents in each opinion cluster is given in the label just above each cluster. Note that all agents in a cluster are connected by positive links, so that interconnected clusters form a clique. Also related, we observe that opinion clusters within a clique are always separated by a distance $\Delta > \varepsilon$.

Third, experimental observation suggest that the BC-PF dynamics with bi-dimensional opinions also produces the segregation of the opinion clusters belonging to different cliques. In particular, in this case, as Fig~\ref{panel-varia-e} illustrates, opinion clusters belonging to different cliques in the absorbing state always get separated by a hyperplane in the issue space. This result is in line with the segregation of opinions in a ``left clique'' and a ``right clique'' that we obtained in the one-dimensional scenario.

\subsection*{Alignment of opinions}

We shall now explore to what extent the phenomena of \textit{unidimensional
opinions} identified by DeMarzo et al~\cite{DeMarzoEA2003} extend to the BC-PF dynamics. DeMarzo et al study a model
of opinion formation \textit{a l\`{a} }Degroot~\cite{DeGroot1974} in weighted networks and show that, in the process of convergence towards consensus, individuals'
opinions over a multidimensional set of issues align in a single
``left-right'' spectrum (the \textit{unidimensional opinions} phenomenon).

In contrast to DeMarzo et al, the BC-PF dynamics does not lead to global consensus of opinions. As we
have seen, opinions in the absorbing state of the BC-PF dynamics differ between agents in different cliques and,
if $\varepsilon $ is small enough, they even differ within agents in the
same clique, which results in the fragmentation of the clique into opinion clusters. As for the alignment of opinions in the BC-PF dynamics, we observe the following. First, that the opinions corresponding to agents in different clusters are not aligned in the opinion space (in the absorbing states). (See, for instance, the three upper panels of Fig~\ref{panel-varia-e} and the bottom-left one, associated to $\varepsilon \leq 0.3$.)
So, in general, we cannot expect an alignment of opinions at the level of the whole population. However, what about opinions at the cluster level?

Fig~\ref{evo-e0p20} sheds some light on this issue. The upper panels in Fig~\ref{evo-e0p20} present a detail of the evolution of agents' opinions associated to the
absorbing state reported in the upper-right panel of Fig~\ref{panel-varia-e} (for $N=20$ and $\varepsilon =0.2$). In
particular, the upper-left panel presents the initial state, the upper-middle panel presents the distribution of opinions after 3 rounds, and the
upper-right panel depicts the absorbing state. In the bottom panels of Fig~\ref{evo-e0p20} we show a zoom of the three
larger opinion clusters of the absorbing state (those containing more than
two agents) after 10 rounds of revisions, i.e., slightly before the
absorbing state is reached.

As we observe in the upper-left panel of Fig~\ref{evo-e0p20}, the initial opinions of the 20 agents are dispersed
around the $[0,1]\times \lbrack 0,1]$ space. The upper-right panel shows
that the system converges to a bipolar state with two cliques containing $9$ and $11$ agents, respectively. Given that $\varepsilon $ is small enough ($\varepsilon=0.2$), in this case each clique gets fragmented in various opinion clusters. In particular, the smaller clique gets partitioned into four opinion
clusters of sizes $1$, $2$, $3$ and $3$; and the larger clique gets
partitioned into two opinion clusters of sizes $1$ and $10$. As already
noted above, the sets of opinion clusters of each one of the two cliques can
be separated by a hyperplane in the opinion space. Attending to the
upper-middle panel, we can observe that after $3$ rounds of revision there
are no two players that share the same opinion, but we can already see how
the opinions of those agents that will end up in the same cluster are
getting closer to each other.

In the bottom-left panel of Fig~\ref{evo-e0p20} we have the zoom corresponding to the opinions of the 10 agents of the
larger cluster (of the larger clique) after 10 rounds. We can observe that,
at this point, the opinions of the 10 agents are almost perfectly aligned.
Likewise, the bottom-midle and bottom-right panels show a zoom corresponding
to each one of the 3-agents clusters of the highest clique. We can also
observe how, within each cluster, the opinions of the three agents are
perfectly aligned after 10 rounds. However, the alignment produced at each
cluster is independent to that produced in others. This result is quite
related to DeMarzo et al, since in the BC-PF dynamics, after a few rounds of evolution, the agents
pioneering an opinion cluster become isolated from the rest of the system
and, from then onwards, our dynamics resembles very much DeMarzo et al's dynamics for fully connected networks with positive unitary links.

Thus, our results say that DeMarzo et al's alignment of opinions
extends to the BC-PF dynamics, but only at the opinion cluster level.

\section*{Potential links to politics}

The results that the BC-PF dynamics produces may have interesting applications to the understating of the political landscape of a country. We discuss that next.

First, note that for $p<1/2$, the BC-PF model always produces the segregation of agents into two polarized cliques with different opinions or, to put
it differently, into two political ideologies, say left and right or liberal and conservative. In this sense, the dynamics of social balance is consistent with the division of the political spectrum in these two clear ideologies.

Second, the BC-PF model can produce both the existence of two cliques with consensus of opinions within cliques, and the existence of two cliques with fragmentation of opinions within cliques. Note that these results hold both in the one-dimensional case and the bi-dimensional case, that is, in cases where agents vote mainly on one issue, say ideology, or when two issues are on campaign, for example ideology and nationalism. (This is usually the case in countries or regions with a strong nationalist feeling, for example Catalonia or Basque Country in Spain, Quebec in Canada, or Flanders in Belgium.) In this sense, variations of our dynamics might prove useful to explain why, at some point in time, a country's traditional two-party system vanishes and a new landscape with two-three political parties in each side of the political spectrum takes the scene. According to our model, it is the agents' confidence level $\varepsilon$ that determines the number of opinion clusters in a clique. (See~\cite{Sobkowicz2016} for a recent study on the emergence of new political parties.)

Taking this kind of results to real world, we might expect a higher political fragmentation after a shock that makes people more vehement and uncompromising, that is, that lowers the value of $\varepsilon$ in the population. Could it be that economic crisis affect people's attitudes towards acquaintances in this sense? Evidence is not clear in this respect, but abundant studies do find that people's trust in institutions (say government, media, business, NGOs, courts of justice, etc.) sharply decreased with the recent economic recession. (See Edelman Trust Barometer or Eurobarometer among others.) In this line, it has also been observed that income inequality, as measured by the Gini index, increased in most OECD countries during the past crisis period and that an increase in income inequality leads to a decrease in trust in people. (See Income Inequality Update 2014, OECD, and Society at a Glance 2011, OECD.) A different class of studies show that agents holding extreme views are less likely to modify their opinions~\cite{BlombergHarrington2000}, and that they are also more likely to: (i) have friends who share their own views on government and politics and (ii) distrust information coming from news sources that do not conform to their political preferences. (See ``Political Polarization and Media Habits'', Pew Research Center October 21, 2014.) According to this, we may expect economic crisis to reduce agents' trust and the latter to be positively related to an agent's confidence level, that is a measure of the agent's willingness to modify his opinion.

Consequently, a possible interpretation of our results would suggest an increase in a country's political fragmentation after an economic crisis (due to the reduction in agents' confidence levels). Interestingly, a recent study by Funke et al~\cite{FunkeEA2016} for 20 advanced economies and more than 800 general elections supports this point. Their study, which covers the period 1870-2014, reveals that after financial crisis government majorities shrink and polarization rises, which increases political uncertainty. The BC-PF model adds to this finding the description of one possible channel through which economic crisis can have a political consequence in terms of a higher political fragmentation. Much has been said in this respect, but most of the arguments talk about economic reasons and the decline in voters' trust in traditional politics and institutions in the aftermath of the financial crash. The BC-PF dynamics could contribute to this discussion showing that the agents' confidence level $\varepsilon$ do play a role in the formation of opinion clusters within cliques, which invites us to think that the argument of the decline of voters' trust in institutions can be complemented with a reference on how economic crisis affect agents' confirmatory bias.

\section*{Discussion and conclusion}

In this work we propose a model, the BC-PF model, that couples in a very natural way Antal et al's~\cite{AntalEA2005} dynamics of structural balance and Hegselman and
Krause's~\cite{HegselmannKrause2002} model of opinions formation under bounded confidence. We rely on two basic ideas: (i) Social relationships change according to a
preferential ``opininion-based'' rule, and (ii) individuals' opinions evolve over time taking into account only the opinions
of their friends in the social network. Because we build on two literatures, the BC$\leftrightarrow $PF model yields a number of results that contribute
to the literature of both structural balance and opinion/beliefs formation. Additionally, the BC$\leftrightarrow $PF model produces results that can be related to real-world phenomena.

Our first and second results refer to the analysis of the BC-PF dynamics under one-dimensional opinions. Regarding social balance and convergence time, we obtain that in line with Antal
et al's LTD model, the BC-PF dynamics also yields convergence to a balanced state -the all-friends configuration- if $p\geq 1/2$, and to a bipolar state otherwise. Similarly, the CTD model in Antal et at also produces convergence to the all-friends configuration for $p\geq 1/2$, whereas for $p<1/2$ the system exhibits a phase transition and it can either be that the all-friends configuration or the bipolar state is reached, depending on the initial friendship density. As for the time to reach the balanced state, it grows with the size of the population much slower in the BC-PF model than in the LTD model. Thus, the BC-PF model provides a new way to reduce the time to reach social balance, allowing us to simulate the dynamics of much larger populations than the LTD model without the need to constrain the dynamics as in the CTD model, or to consider a continuous-time dynamics as in~\cite{KulakowskiEA2005,MarvelEA2011,SrinivasanEA2011}.

Our second result regards to the segregation of opinions. Focusing on the
more interesting case of $p<1/2$, when $\varepsilon $ is large enough,
the BC-PF model leads to consensus of opinions within cliques, i.e., two opinions survive in the long run, one for each
clique. Instead, for lower values
of $\varepsilon $, in the absorbing state of the BC-PF
model, more than one opinion can coexist within each clique. Hence, our model extends Hegselmann and
Krause's main result (the existence of a number of opinion clusters
that inversely depend on $\varepsilon $, which collapses into global
consensus for $\varepsilon $ large enough) to the case in which the
population segregates into two cliques. Interestingly, the BC$%
\leftrightarrow $PF dynamics segregates the agents into the two cliques
according to their opinions: The opinion clusters in one clique are all
below (or above) the opinion clusters in the other clique, i.e., it yields a ``left clique'' and a ``right clique''.

Our third and fourth results refer to the analysis of the BC-PF dynamics under bi-dimensional opinions. In this
respect, Fortunato et al~\cite{FortunatoEA2005} study the bi-dimensional extension of the
model by Hegselmann and Krause and find that, starting from uniform
probability distribution for the opinion configuration, when $\varepsilon $
is large enough, consensus is reached whereas, for smaller $\varepsilon $,
opinion clusters arise and typically form a lattice in the opinions space. In line with the one-dimensional case, we find that the BC-PF model also yields segregation of agents into two
cliques for $p<1/2$, being the number of opinions within each clique inversely
related to $\varepsilon $. If $\varepsilon $ is large enough, all agents
within a clique share the same opinion (a bi-dimensional vector), which
differs from the opinion of the agents in the other clique. For lower $%
\varepsilon $, different opinion clusters coexist within each clique.
Interestingly, in all our simulations the set of opinion clusters in the first clique and that in the second clique can be separated by a hyperplane in the opinions space, which extends the result of the segregation in a ``left clique'' and
a ``right clique'' that we obtain in the one-dimensional scenario.

Our final result refers to a phenomenon identified by DeMarzo et al~\cite{DeMarzoEA2003},
who study the evolution of multidimensional opinions \textit{\`{a} la}
DeGroot~\cite{DeGroot1974}. Their main result is that, at some point in the convergence process towards consensus, all the agents' opinions align and keep aligned thereafter. Thus, they claim that the individuals' opinions
over a multidimensional set of issues converge to a single ``left-right'' spectrum (a phenomenon that they label as unidimensional opinions). We find that this result extends partially to the bi-dimensional version of our BC-PF model. In particular, we obtain that the opinions of
all the agents within each cluster do align at some point of the convergence process. However, this alignment only holds at the level of opinion clusters, and not at the population level. Hence, our results suggest that for dynamics not driving to global consensus, the main result of DeMarzo et al would be expected to hold partially.

Last, we discuss some implications of our results to politics. We observe that the BC-PF model can produce the segregation of agents into two polarized cliques with both consensus of opinions within a clique or fragmentation of opinions within it. Interestingly, these results have a very natural counterpart in the political landscape of a country. Indeed, there are countries -or time periods- where the traditional two-party system dominates, showing a (possibly just superficial) consensus of opinions within each side of the political spectrum;  whereas in other periods of time all of a sudden a side of the political spectrum fragments and new parties arise. The BC-PF model invites us to think about this phenomena in terms of agents' confidence level, confirmatory bias and, more broadly, in terms of agents' trust. Recent data from Edelman Trust Barometer and Eurobarometer shows that people's trust decline with economic crisis. The study by Funke et al~\cite{FunkeEA2016} show that economic crisis cause political fragmentation. Interestingly, the results in the BC-PF model suggest a possible link between these two lines of research, talking about agents' confidence levels or trust as a nexus that could connect economic variables with political variables.

Finally, we believe that this paper is only one of the first steps in a
broader research agenda, which deals with the study of the joint evolution
of opinions/beliefs and social ties. Indeed, the studies dealing with the joint evolution of agents' opinions and networks are still relatively scarce. (See the Related Literature section.) A clearer understanding of the
effects arising from the complexity of this kind of (intertwined) dynamics
may well be the result of the knowledge gleaned from these and future models. In this respect, it would be interesting to study alternative
dynamics, for instance the opinion formation model proposed
by Deffuant et al~\cite{DeffuantEA2000}. Other extension could be to introduce repulsion to enemies' opinions, as considered for instance, by~\cite{Altafini2012,Altafini2013,FlacheMacy2011,FanEA2012,XiaEA2016,ProskurnikovEA2016,MengEA2016,ShiEA2016} in the context of fixed
networks. We conjecture that this would increase polarization which, in the
bi-dimensional case, is likely to yield the separation of the opinion
clusters of the two cliques in both issues (i.e., a separation in
two diagonal quadrants of the opinion space). Another (complementary)
extension would be to consider incomplete networks. As for the field of applications, despite the difficulties inherent in obtaining direct evidence of social influence (see Parravano et al~\cite{ParravanoEA2015} for an example), it would be interesting to further explore connections of this type of models to real world phenomena. These, as well as
other possible extensions of the model, are left for future research.

\section*{Acknowledgments}
We thank an anonymous referee for helpful comments. We gratefully acknowledge financial support from the Ministerio de Educaci\'{o}n y Ciencia through project ECO2011-26996, and the Junta de Andaluc\'{\i}a through project SEJ2011-8065.

\nolinenumbers

%
%
%


\begin{thebibliography}{10}

\bibitem{Byrne71}
Byrne D. The attraction paradigm. New York: Academic Press; 1971.

\bibitem{Kandel78}
Kandel DB. Homophily, selection, and socialization in adolescent
friendships. American Journal of Sociology. 1978; 84:427--436.

\bibitem{Mark2003}
Mark NP. Culture and competition: homophily and distancing
explanations for cultural niches. American Sociological Review. 2003; 68(3):319--345.

\bibitem{ParravanoEA2007}
Parravano A, Rivera-Ram\'{\i}rez H, Cosenza MG. Intracultural diversity in a model of social dynamics. Physica A. 2007; 379:241--249.

\bibitem{CurrariniEA2009}
Currarini S, Jackson MO, Pin P. An economic model of
friendship: homophily, minorities and segregation. Econometrica 2009; 77:1003--1045.

\bibitem{CurrariniEA2010}
Currarini S, Jackson MO, Pin P. Identifying the roles of
race-based choice and chance in high school friendship network formation.
Proceedings of the National Academy of Sciences. 2010; 107(11):4857--4861.

\bibitem{FlacheMacy2011b}
Flache A, Macy MW. Local convergence and global diversity:
from interpersonal to social influence. Journal of Conflict Resolution. 2011; 55(6):970--995.

\bibitem{McPhersonEA2001}
McPherson M, Smith-Lovin L, Cook JM. Birds of a feather:
homophily in social networks. Annual Review of Sociology. 2001; 27:415--444.

\bibitem{AntalEA2005}
Antal T, Krapivsky PL, Redner S. Dynamics of social balance
networks. Physical Review E. 2005; 72, 036121.

\bibitem{Heider1946}
Heider F. Attitudes and cognitive organization. Journal of
Psychology. 1946; 21:107--112.

\bibitem{CartwrightHarary1956}
Cartwright D, Harary F. Structural balance: a generalization
of Heider's theory. Psychological Review. 1956; 63(5):277--293.

\bibitem{FacchettiEA2011}
Facchetti G, Iacono G, Altafini C. Computing global
structural balance in large-scale signed social networks. Proceedings of the
National Academy of Sciences. 2011; 108(52):20953--20958.

\bibitem{HegselmannKrause2002}
Hegselmann R, Krause U. Opinion dynamics and bounded confidence
models, analysis, and simulation. Journal of Artificial Societies and Social
Simulation. 2002; 5(3).

\bibitem{LordEA1979}
Lord CG, Ross L, Lepper MR. Biased assimilation and
attitude polarization: The effects of prior theories on subsequently
considered evidence. Journal of Personality and Social Psychology. 1979; 37(11):2098--2109.

\bibitem{LazarsfeldEA1948}
Lazarsfeld PF, Berelson B, Gaudet H. The people's choice.
New York: Columbia University Press; 1948.

\bibitem{LazarsfeldEA1954}
Lazarsfeld PF, Lipset SM, Barton AH, Linz J. The
psychology of voting: an analysis of political behavior. In: G. Lindzey
(Ed.), Handbook of social psychology, Vol II. Cambridge: Addison-Wesley; 1954.

\bibitem{Andina2007}
Andina-D\'{\i}az A. Reinforcement vs. change: the political
influence of the media. Public Choice. 2007; 131:65--81.

\bibitem{Allahverdyan2014}
Allahverdyan AE, Galstyan A. Opinion dynamics with confirmation bias. PLOS ONE. 2014; 9(7):e99557.

\bibitem{BessiEA2015}
Bessi A, Coletto M, Davidescu GA, Scala A, Caldarelli G, Quattrociocchi W. Science vs conspiracy: collective narratives in the
age of misinformation. PLOS ONE. 2015; 10(2):e0118093.

\bibitem{DelVicarioEA2016a}
Del Vicario M, Scala A, Caldarelli G, Stanley HE, Quattrociocchi W. Modeling confirmation bias and polarization. 2016; arXiv:1607.00022.

\bibitem{DelVicarioEA2016b}
Del Vicario M, Bessi A, Zollo F, Petroni F, Scala A, Caldarelli G, et al. The spreading of misinformation
online. Proceedings of the National Academy of Sciences. 2016; 113(3):554--559.

\bibitem{DeGroot1974}
DeGroot MH. Reaching a consensus. Journal of the American
Statistical Association. 1974; 69(345):118--121.

\bibitem{DeMarzoEA2003}
DeMarzo PM, Vayanos D, Zwiebel J. Persuasion bias, social influence, and unidimensional opinions. Quarterly Journal of
Economics. 2003; 118(3):909--968.

\bibitem{AntalEA2006}
Antal T, Krapivsky PL, Redner S. Social balance on
networks: the dynamics of friendship and enmity. Physica D: Nonlinear
Phenomena. 2006; 224:130--136.

\bibitem{ZhengEA2015}
Zheng X, Zeng D, Wang FY. Social balance in signed
networks. Information Systems Frontiers. 2015; 17(5):1077--1095.

\bibitem{KulakowskiEA2005}
Ku\l akowski K, Gawronski P, Gronek P. The Heider balance --
a continuous approach. The International Journal of Modern Physics C. 2005; 16:707.

\bibitem{MarvelEA2011}
Marvel S, Kleinbergb JJ, Robert D, Kleinberg RD, Strogatz S. Continuous-time model of structural balance. Proceedings of the
National Academy of Sciences. 2011; 108:1771-1776.

\bibitem{SrinivasanEA2011}
Srinivasan A. Local balancing influences global structure in social
networks. Proceedings of the National Academy of Sciences. 2011; 108(5):1751--1752.

\bibitem{TraagEA2013}
Traag VA, Van Dooren P, De Leenheer P. Dynamical models
explaining social balance and evolution of cooperation. PLOS ONE. 2013; 8(4):e60063.

\bibitem{AguiarParravano2015}
Aguiar F, Parravano A. Tolerating the intolerant: homophily,
intolerance, and segregation in social balanced networks. Journal of
Conflict Resolution. 2015; 59(1):29--50.

\bibitem{French1956}
French JR. A formal theory of social power. Psychological
Review. 1956; 63(3):181--194.

\bibitem{Harary1959}
Harary F. A criterion for unanimity in French's theory of social
power. In Cartwright D. (Ed.), Studies in Social Power. Institute for Social
Research, Ann Arbor; 1959.

\bibitem{FriedkinEA1990}
Friedkin NE, Johnsen EC. Social influence and opinions.
Journal of Mathematical Sociology. 1990; 15(3-4):193--205.

\bibitem{FriedkinEA1999}
Friedkin NE, Johnsen EC. Social influence networks and
opinion change. Advances in Group Processes. 1999; 16:1--29.

\bibitem{Golub2010}
Golub B, Jackson MO. Na\"{\i}ve learning in social networks
and the wisdom of crowds. American Economic Journal: Microeconomics. 2010; 2(1):112--149.

\bibitem{DeffuantEA2000}
Deffuant G, Neau D, Amblard F, Weisbuch G. Mixing beliefs
among interacting agents. Advances in Complex Systems. 2000; 3(01n04):87--98.

\bibitem{FortunatoEA2005}
Fortunato S, Latora V, Pluchino A, Rapisarda A. Vector
opinion dynamics in a bounded confidence consensus model. International
Journal of Modern Physics C. 2005; 16(10):1535--1551.

\bibitem{PluchinoEA2006}
Pluchino A, Latora V, Rapisarda A. Compromise and
synchronization in opinion dynamics. European Physical Journal B. 2006; 50:169--176.

\bibitem{FlacheMacy2011}
Flache A, Macy MW. Small worlds and cultural polarization.
The Journal of Mathematical Sociology. 2011; 35(1-3):146--117.

\bibitem{KouEA2012}
Kou G, Zhao Y, Peng Y, Shi Y. Multi-level opinion dynamics
under bounded confidence. PLOS ONE. 2012; 7(9):e43507.

\bibitem{FuEA2015}
Fu G, Zhang W, Li Z. Opinion dynamics of modified
Hegselmann--Krause model in a group-based population with heterogeneous
bounded confidence. Physica A. 2015; 419:558--565.

\bibitem{LorenzEA2007}
Lorenz J. Continuous opinion dynamics under bounded confidence: A
survey. International Journal of Modern Physics C. 2007; 18(12):1819--1838.

\bibitem{CastellanoEA2009}
Castellano C, Fortunato S, Loreto V. Statistical physics of
social dynamics. Reviews of Modern Physics. 2009; 81(2):591.

\bibitem{AlbiEA2016}
Albi G, Pareschi L, Toscani G, Zanella M. Recent advances in
opinion modeling: control and social influence. 2016; arXiv:1607.05853.

\bibitem{DandekaraEA2013}
Dandekar P, Goel A, Lee DT. Biased assimilation,
homophily, and the dynamics of polarization. Proceedings of the National
Academy of Sciences. 2013; 110(15):5791--5796.

\bibitem{GroeberEA2014}
Groeber P, Lorenz J, Schweitzer F. Dissonance minimization as
a microfoundation of social influence in models of opinion formation. The
Journal of Mathematical Sociology. 2014; 38(3):147--174

\bibitem{Altafini2012}
Altafini C. Dynamics of opinion forming in structurally balanced
social networks. PLOS ONE. 2012; 7(6):e38135.

\bibitem{Altafini2013}
Altafini C. Consensus problems on networks with antagonistic
interactions. IEEE Transactions on Automatic Control. 2013; 58(4):935--946.

\bibitem{HuEA2014}
Hu J, Zheng WX. Emergent collective behaviors on coopetition
networks. Physics Letters A. 2014; 378(26-27):1787--1796.

\bibitem{FanEA2012}
Fan P, Wang H, Li P, Li P, Jiang Z. Analysis of opinion
spreading in homogeneous networks with signed relationships. Journal of
Statistical Mechanics: Theory and Experiment. 2012; P08003.

\bibitem{XiaEA2016}
Xia W, Cao M, Johansson KH. Structural balance and opinion
separation in trust-mistrust social networks. IEEE Transactions on Control
of Network Systems. 2016; 3(1):46--56.

\bibitem{ProskurnikovEA2016}
Proskurnikov AV, Matveev A, Cao M. Opinion dynamics in
social networks with hostile camps: Consensus vs. polarization. IEEE
Transactions on Automatic Control. 2016; 61(6):1524--1536.

\bibitem{MengEA2016}
Meng Z, Shi G, Johansson KH, Cao M, Hong Y. Behaviors of
networks with antagonistic interactions and switching topologies. 2016; arXiv:1402.2766.

\bibitem{ShiEA2016}
Shi G, Proutiere A, Johansson M, Baras JS, Johansson KH. The evolution of beliefs over signed social networks. Operations Research. 2016; 64(3):585--604.

\bibitem{MasFlache2013}
M\"{a}s M, Flache A. Differentiation without distancing.
Explaining bi-polarization of opinions without negative influence. PLOS ONE. 2013; 8(11):e74516.

\bibitem{HolmeEA2006}
Holme P, Newman MEJ. Nonequilibrium phase transition in the coevolution of networks and opinions. Physical Review E. 2006; 74, 056108.

\bibitem{FuWang2008}
Fu F, Wang L. Coevolutionary dynamics of opinions and networks: from diverity to uniformity. Physical Review E. 2008; 78, 016104.

\bibitem{NardiniEA2008}
Nardini C, Kozma B, Barrat A. Who’s talking first? Consensus or lack thereof in coevolving opinion formation models. Physical Review Letters. 2008; 100, 158701.

\bibitem{VazquezAE2008}
Vazquez F, Eguíluz VM, San Miguel M. Generic Absorbing Transition in Coevolution Dynamics. Physical Review Letters. 2008; 100, 108702.

\bibitem{IniguezEA2009}
Iñiguez G, Kertész J, Kaski KK, Barrio RA. Opinion and community formation in coevolving networks. Physical Review E. 2009; 80, 066119.

\bibitem{DurrettEA2012}
Durrett R, Gleeson JP, Lloyd AL, Mucha PJ, Shi F, Sivakoff D, et al. Graph fission in an evolving voter model. Proceedings of the National Academy of Sciences. 2012; 109(10):3682--3687.

\bibitem{YiEA2013}
Yi SD, Baek SK, Zhu CP, Kim BJ. Phase transition in a coevolving network of conformist and contrarian voters. Physical Review E. 2013; 87, 012806.

\bibitem{ArifovicEA2015}
Arifovic J, Eaton BC, Walker G. The coevolution of beliefs and networks. Journal of Economic Behavior and Organization. 2015; 120:46--63.

\bibitem{KozmaBarrat2008}
Kozma B, Barrat A. Consensus formation on adaptive networks. Physical Review E. 2008; 77, 016102.

\bibitem{GilZanette2006}
Gil S, Zanette DH. Coevolution of agents and networks: opinion spreading and community disconnection. Physics Letters A. 2006; 356:89--94.

\bibitem{ZanetteGil2006}
Zanette DH, Gil S. Opinion spreading and agent segregation on evolving networks. Physica D. 2006; 224:156--165.

\bibitem{EhrhardtEA2006}
Ehrhardt G, Marsili M, Vega-Redondo F. Diffusion and growth in an evolving network. International Journal of Game Theory. 2006; 34:383--397.

\bibitem{EhrhardtEA2008}
Ehrhardt G, Marsili M, Vega-Redondo F. Emergence and resilience of social networks: a general theoretical framework. Annales d'Économie et de Statistique. 2008; N 86.

\bibitem{JacksonWatts2002}
Jackson MO, Watts A. On the formation  of interaction  networks  in social  coordination  games. Games and Economic Behavior. 2002; 41(2):265--291.

\bibitem{GoyalVega2005}
Goyal S, Vega-Redondo F. Learning, network formation, and coordination, Games and Economic Behavior. 2005; 50:178--207.

\bibitem{FeriMelendez2013}
Feri F, Meléndez-Jiménez MA. Coordination in evolving networks with endogenous decay. Journal of Evolutionary Economics. 2013; 23:955--1000.

\bibitem{HellmanStaudigl2014}
Hellmann T, Staudigl M. Evolution of social networks. European Journal of Operational Research. 2014; 234(3):583--596.

\bibitem{BenczikEA2009}
Benczik IJ, Benczik SZ, Schmittmann B, Zia RKP. Opinion dynamics on an adaptive random network. Physical Review E. 2009; 79, 046104.

\bibitem{Abel2009}
Abell P, Ludwig M. Structural balance: a dynamic perspective.
The Journal of Mathematical Sociology. 2009; 33(2):129--155.

\bibitem{Sobkowicz2016}
Sobkowicz P. Quantitative agent based model of opinion dynamics:
Polish elections of 2015. PLOS ONE. 2016; 11(5):e0155098.

\bibitem{BlombergHarrington2000}
Blomberg SB, Harrington JE. A theory of rigid extremist and flecible moderates with an application to the {U.S.} {C}ongress. American Economic Reivew. 2000; 90(3):605--620.

\bibitem{FunkeEA2016}
Funke M, Schularick M, Trebesch C. Going to extremes: politics after financial crises, 1870–2014. European Economic Review. 2016; 88:227--260.

\bibitem{ParravanoEA2015}
Parravano A, Noguera JA, Hermida P, Tena-S\'{a}nchez J. Field evidence of social influence in the expression of political
preferences: the case of secessionists flags in Barcelona. PLOS ONE. 2015; 10(5):e0125085.

\end{thebibliography}
\end{document}